\documentclass{aa}
\usepackage{graphicx}
\usepackage{txfonts}
\begin{document}

\title{Revisiting Pushchino RRAT search using neural network}
   
   \author{S.A. Tyul'bashev\inst{1}
          \and
          D.V. Pervukhin\inst{1}
          \and
          M.A. Kitaeva\inst{1}
          \and
          G.E. Tyul'basheva\inst{2}
          \and
          E.A. Brylyakova\inst{1}
          \and
          A.V. Chernosov\inst{3}
          }

  \institute{Lebedev Physical Institute, Astro Space Center, Pushchino Radio Astronomy Observatory\\
142290, Radioteleskopnaya 1a, Moscow reg., Pushchino, Russia\\
  \email{serg@prao.ru}
                     \and
 Institute of Mathematical Problems of Biology, brunch of Keldysh Institute of Applied Mathematics\\
142290, Vitkevich 1, Moscow reg., Pushchino, Russia
                     \and
The State University of Management\\ 
102542, Ryazanskii prospect 99, Moscow, Russia
             }

   \date{Received ; accepted}
   
\abstract
{The search for rotating radio transients (RRAT) at declination from -9$^o$ to +42$^o$ was carried out in the semi-annual monitoring data obtained on the Large Phased Array (LPA) radio telescope at the frequency of 111~MHz. A neural network was used to search for candidates. 4 new RRATs were detected, having dispersion measures (DM) 5-16 pc/cm$^3$. A comparison with an earlier RRAT search conducted using the same data shows that the neural network reduced the amount of interference by 80 times, down to 1.3\% of the initial amount of interferences. The loss of real pulsar pulses does not exceed 6\% of their total number.}


   \maketitle
%

\section{Introduction}

Rotating radio transients (RRAT), being pulsars with sporadic radiation, were discovered 15 years ago (\citeauthor{McLaughlin2006}, \citeyear{McLaughlin2006}). The average time intervals between the emitted pulses start from minutes and can reach up to many hours (\citeauthor{McLaughlin2006}, \citeyear{McLaughlin2006}, \citeauthor{Logvinenko2020}, \citeyear{Logvinenko2020})). Despite the actively ongoing search for new pulsars, including the search for RRATs (\citeauthor{Deneva2016} (\citeyear{Deneva2016}), \citeauthor{Tyulbashev2018b} (\citeyear{Tyulbashev2018b}), \citeauthor{Sanidas2019} (\citeyear{Sanidas2019}), \citeauthor{Han2021} (\citeyear{Han2021}), \citeauthor{Good2021} (\citeyear{Good2021})), as well as  re-processing of archived data (\citeauthor{Keane2010} (\citeyear{Keane2010}), \citeauthor{Burke-Spolaor2010} (\citeyear{Burke-Spolaor2010}), \citeauthor{Keane2011} (\citeyear{Keane2011}), \citeauthor{Karako-Argaman2015} (\citeyear{Karako-Argaman2015})), in ATNF catalog\footnote{$https://www.atnf.csiro.au/people/pulsar/psrcat/$} (\citeauthor{Manchester2005}, \citeyear{Manchester2005}) as of November 2021, there are only 109 RRATs. 

There is no unambiguous understanding yet of what RRAT is and what the reasons for such a rare appearance of its impulses are. There are a number of phenomenological hypotheses about the nature of RRAT. The four simplest hypotheses suggest that RRATs are ordinary pulsars (\citeauthor{Weltevrede2006} (\citeyear{Weltevrede2006}), \citeauthor{Zhang2007} (\citeyear{Zhang2007}), \citeauthor{Wang2007} (\citeyear{Wang2007}), \citeauthor{Brylyakova2021} (\citeyear{Brylyakova2021})). These hypotheses are based on the properties of the detected pulses: the distribution of pulses by energy, frequency of pulses occurrence, the observed flux density and other objective indicators. For example, RRATs can be ordinary pulsars with a long tails of pulse energy distribution (\citeauthor{Weltevrede2006}, \citeyear{Weltevrede2006}), or RRATs can be pulsars with very long nullings (\citeauthor{Zhang2007}, \citeyear{Zhang2007}), or RRATs are pulsars with an extreme case of switching modes  (\citeauthor{Wang2007}, \citeyear{Wang2007}), or RRATs may be pulsars with giant pulses (\citeauthor{Brylyakova2021} (\citeyear{Brylyakova2021}), \citeauthor{Tyulbashev2021} (\citeyear{Tyulbashev2021})). In the paper \citeauthor{Burke-Spolaor2010} (\citeyear{Burke-Spolaor2010}) there was an assumption that there is some kind of ''uncompromising`` version of the RRAT, but not a single example of such a RRAT was given.

RRAT search is hampered. The field of view of full-rotation or stationary spherical/parabolic mirrors is small, and there is also little time allocated to viewing one direction in the sky. For example, in surveys conducted on radio telescopes (RT) RT-64 (Parks), RT-100 (Effelsberg and Green-Bank), RT-300 (Aresibo), RT-500 (FAST) (\citeauthor{Lorimer2006} (\citeyear{Lorimer2006}), \citeauthor{Deneva2009} (\citeyear{Deneva2009}), \citeauthor{Barr2013} (\citeyear{Barr2013}), \citeauthor{Boyles2013} (\citeyear{Boyles2013}),  \citeauthor{Han2021} (\citeyear{Han2021})), the typical observation session time for the selected direction is a minute or several minutes, and therefore transients whose intervals between successive pulses are much longer than a few minutes will be skipped. The sensitivity of these radio telescopes is high, since in the decimeter wavelength range (typically 327-1400 MHz), where the search is carried out, the background temperature of the Galaxy is low, and the frequency bands used for observations are wide. Therefore, pulses similar to those of ordinary strong pulsars are recorded confidently.

Separately, we note the CHIME radio telescope. This radio telescope operating in the decimeter range has pulsars and RRATs detected\footnote{$https://www.chime-frb.ca/galactic$} (\citeauthor{Good2021}, \citeyear{Good2021}). They are a by-products when searching for fast radio bursts (FRBs). However, a wide field of view, an effective area comparable to the area of hundred-meter mirrors, and a wide reception band suggest its likely high efficiency in organizing a targeted search for RRAT pulsars.

Antenna arrays of meter wavelength range used to search for pulsars, for example, LPA (16,384 dipoles (\citeauthor{Tyulbashev2016}, \citeyear{Tyulbashev2016}, \citeauthor{Tyulbashev2018a}, \citeyear{Tyulbashev2018a})) and LOFAR (central core of 4096 dipoles  (\citeauthor{Sanidas2019}, \citeyear{Sanidas2019})) can have a field of view of up to several dozens of square degrees. RRAT search surveys on these telescopes should be more time-efficient than on spherical or parabolic mirrors. However, when searching for RRAT, the main parameter is the instantaneous sensitivity of the antenna, and it is determined primarily by the effective area and the Galaxy background temperature. Unfortunately, the Galaxy background temperature is high in the meter wavelength range, and therefore the listed arrays can be real competitors of conventional mirrors operating in the decimeter wavelength range - either due to a large effective area (LPA antenna) or due to a wide reception band (LOFAR antenna).

Taking into account the large time intervals between successive pulses, dozens of observation hours of each point in the sky are needed when searching for RRAT, and it is not surprising that about half of RRATs were found during the reprocessing of archived data, and not during the targeted search. For example, when reprocessing GBT data, 21 RRATs were found (\citeauthor{Karako-Argaman2015}, \citeyear{Karako-Argaman2015}), and when reprocessing the Parks telescope data, 28 RRATs were found (\citeauthor{McLaughlin2006} (\citeyear{McLaughlin2006}), \citeauthor{Keane2010} (\citeyear{Keane2010}), \citeauthor{Burke-Spolaor2010} (\citeyear{Burke-Spolaor2010})). 

In addition to the longer-time intervals required when searching for RRAT, search algorithms that guarantee high reliability of detecting new objects are also very important. So, in the work on searching RRAT using 48 LPA beams covering declinations $+21^o<\delta<+42^o$, and processing of one month of round-the-clock monitoring observations, $10^7$ candidates were found \citeauthor{Tyulbashev2018a} (\citeyear{Tyulbashev2018a}). As noted in this paper, different digital filters reduced the number of candidates to $10^5$, and subsequent visual inspection allowed the selection of new RRATs. When searching for RRAT in monitoring data for an interval of 6 months, the number of candidates increased up to $6\times 10^5$, and when additional digital filters were involved, it decreased down to $3\times 10^5$, which were visually checked (\citeauthor{Tyulbashev2018b}, \citeyear{Tyulbashev2018b}). 
Currently, the daily monitoring survey has been going on for 7.5 years and, therefore, the expected number of candidates being tested may increase up to $5\times 10^6$. It is not realistic to visually check such a number of candidates. In this paper, we consider the possibility of using a recurrent neural network (RNN) to identify reliable candidates and filter out false sources.

\section{Using a neural network to select sources}


In the paper \citeauthor{Tyulbashev2018a} (\citeyear{Tyulbashev2018a}), it was said that the initial search for dispersed pulses was carried out in a standard way: a) dispersion measures ($DM$) were sorted out from 3 to 100 pc/cm$^3$; b) frequency channels were added up for each sorted out $DM$, the baselines were subtracted; c) the standard deviations of noise ($\sigma_n$) were estimated; d) the points of the array obtained from the sorted out $DM$ were checked sequentially. If for some point the signal-to-noise ratio ($S/N=A/\sigma_n$, where $A$ is a signal amplitude) is higher than five on this $DM$, the picture of the dynamic spectrum and the added-up profile was memorized; e) during visual inspection, interferences were eliminated, and candidates for RRAT were additionally checked if at least one pulse having $S/N\ge 7$ was detected.

In the processing method used, strong pulses will be recorded at different $DM$ as long as the $S/N$ of the assembled profile remains higher than the set level. With $DM$ close to actual, a profile amplitude will be higher, and a profile width will be less. The higher the observed $S/N$ ratio for the pulse, the more times it will be detected on different $DM$, and the more pictures of the same pulse will be memorized.

When viewing one image of the dynamic spectrum per one second, the full viewing time of the semi-annual processing of observations on the RRAT search ($6\times 10^5$ images), will take 600 thousand seconds or, approximately, one working month. In practice, the initial viewing of images was done quickly. The processing program can sequentially display all dynamic spectra for a given direction in the sky in ascending order of $DM$, or for a given sidereal time in an interval of about 3.5 minutes, sequentially going through all the beams, and in each beam, in ascending order, all $DM$. When we quickly flip through the dynamic spectra, inclined lines are visible on the screen, indicating signals with a dispersion delay. 
If we notice an inclined line when viewing, then we return to this picture (or pictures) and evaluate the characteristics of the pulse (coordinates, $DM$). These characteristics allow us to make a subsequent identification of the source in ATNF. For clarity, we give an example of processing in dispersed pulses search process\footnote{$http://prao.ru/online\%20data/onlinedata.html$}. 


In the paper \citeauthor{Tyulbashev2018b} (\citeyear{Tyulbashev2018b}), it was said that when processing semi-annual data, about a quarter of all dynamic spectra indicated real pulsar pulses and RRATs, and three quarters of the detected pulses were associated with interferences. Some of these interferences are similar to the pulses of ordinary pulsars having $DM$, as a rule, in the range from 3 to 10 pc/cm$^3$.  The elimination of such interferences is based on the fact that these interference signals can be observed simultaneously in many beams, or, during the day, they are observed at very different right ascensions beyond the dimensions of the receiving beam of the LPA of Lebedev Physical Institute (LPI). The total transit time of the source through the meridian is approximately seven minutes. Therefore, if the pulses with the same $DM$  are located on a given day far from each other in time ($>$7 minutes), then we can talk about detecting an interference. Some other types of interference are discussed below.
\begin{figure*}
   \centering
   \includegraphics{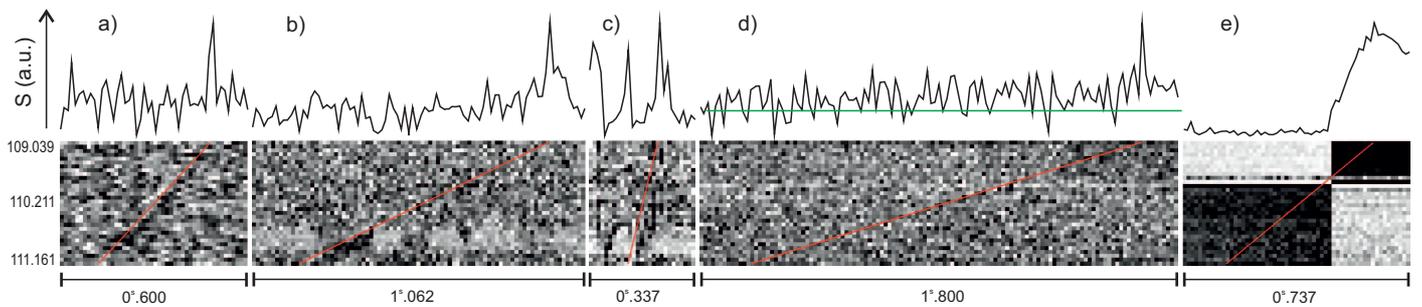}
   \caption{Examples of interferences, missed by the neural network. The lower part of the figure is the dynamic spectra, the red line is drawn along the candidate RRAT found, and the degree of the line slope reflects $DM$. Above the dynamic spectrum, a profile is shown, assembled taking into account the expected $DM$. The recording duration in seconds is shown under each dynamic spectrum. The frequencies of some channels are shown vertically near the dynamic spectrum, and near the profile, there is a flux density in conventional units. Figure 1d shows in green the baseline from which the signal amplitude was calculated.}
    \end{figure*} 
Neural networks can be used to reduce the number of interferences in visually verified images. For example, convolutional neural networks (CNN), previously used to analyze data obtained from LPA LPI, have shown high efficiency (\citeauthor{Isaev2019}, \citeyear{Isaev2019}). When processing data using the CNN neural network model, the time series was divided into fixed-length frames, which made it possible to use CNN, which works well when searching for objects in images.

In this paper, in the pictures of dynamic spectra, a found transient is a line of dark pixels starting in frequency channels at a high frequency and shifting in time to a lower frequency. The magnitude of the line slope is related to $DM$. That is, from the point of view of pattern recognition, we have simple images of lines at different slopes, ''diagonals``, that the network should ''see``. The length of this diagonal is the higher, the larger is $DM$, and this means that we must be able to process frames of different lengths with the network. However, on frames of different lengths, there may be a sharp decrease in the quality of CNN learning, and, consequently, in recognition.

The neural network model is written in Python and based on a Google Tensorflow\footnote{$https://www.tensorflow.org/api\_docs/python/tf$} end-to-end open source platform for machine learning. Pandas and Numpy libraries were used for a basic ETL (Extract-Transform-Load) processing, Keras - as a more generalised approach to design and train a particular neural network model for the classification problem. Besides, a distributed cloud web system was designed on Python/MongoDB stack for ETL and visualisation.

A core architecture of the model is an LSTM (Long-Short-Term Memory)  (\citeauthor{Hochreiter1997}, \citeyear{Hochreiter1997}), which is, in turn, based on a concept of artificial RNN, oftenly used in the field of deep learning (DL). To train the model a common approach was used. The raw data was converted to a unified format during the ETL procedure, and combined into Pandas dataframes to speedup access and optimize storage. 

The trained model accepts a tensor in a unified format and converts it into a resulting value, ranging from 0.00 (probably not positive) to 1.00 (probably positive). To make a resulting value, which is boolean by its nature, we need to find out a threshold, when model values below are interpreted as a false when values above are interpreted as a true. So for the final tuning of the model the threshold is estimated to minimise a number of type I and type II errors on the test dataset. Final model testing was made on real production data with manual checks of the model execution results.


For the network train, we have taken approximately 9,000 images from $6\times 10^5$ of them obtained during the processing of semi-annual observations, and entered the following type information into the database: ''YES`` means we see a transient in the image; ''NO`` means we see an interference. There were about equal numbers of ''YES`` and ''NO`` images. These images were divided in the ratio of 2/3 and 1/3, where 2/3 of the images were used for network training (training data), and 1/3 of them was used to test the network after training (test data).


After receiving the model, different thresholds for making the right decision were obtained based on the test data, these thresholds were tested on all the processed data. Testing of these thresholds showed that at the level of the threshold for making the right decision equal to 0.8 and with an acceptable number of remaining interferences, there is a small number of missed transients (some details is below in this section).

After the end of the network training process, the sample images used for training were deleted, and the search for transients was carried out throughout the remaining samples of candidates. Since in the new search we used previously processed data  (\citeauthor{Tyulbashev2018b}, \citeyear{Tyulbashev2018b}), namely, images of dynamic spectra and pulse profiles, for each direction in the sky, the number of observed interferences and the number of detected pulses were already known. Consequently, after the operation of the neural network, it was possible to view the remaining images, and check how many interferences the network took for real pulses, and how many real pulses the network threw out, considering them as interferences. In other words, it is possible to estimate in practice the probability of a false positive detection (a second-kind error) and the probability of missing a real signal (a first-kind error). 

When processing observation data, we divide the entire sky into pixels, the size of which is slightly smaller than the size of the LPA LPI receiving beam, and perform an independent search in each pixel. Since the exact coordinate of a transient is not known, its pulses can be detected in several nearby pixels by right ascension and declination. For used configuration of LPA, the entire available sky consists of 96 beams divided by declination ($-9^o<\delta <+42^o$), with a distance between the beams of about $0.5^o$, and located in the meridian plane, as well as from 422 pieces with a duration of approximately 3.5 minutes by right ascension: $(3.5\times 422)/60=24$~h. Thus, the sky is divided into $96\times 422 = 40,512$ pixels.

After the neural network operation, it turned out that for 97\% of all pixels, on average, less than one interference record per pixel is detected. Almost half of all pixels has not a single image of the dynamic spectrum left. The median value of the remaining interferences number per 96 pixels (corresponding to 96 beams in the meridian) is 12.

Three percent of the recordings show from dozens to hundreds of interference recordings per pixel. It turned out that we missed one type of interference when we trained the neural network, and it takes these interferences for transients. We were not able to find out a nature of these interferences, but we hope to learn how to eliminate them with further improvements of the neural network when processing data accumulated over an interval of 7 years. In Fig.1e, we present the dynamic spectrum and the average profile for such interferences. Excluding 3\% of pixels from the analysis, we get that the neural network removed almost 99\% of all interferences that the original processing program took for dispersed pulses.

Another important question is how many real pulses are removed by the neural network. To estimate the number of pulses removed by the network from detected during early processing of the semi-annual data (\citeauthor{Tyulbashev2018b}, \citeyear{Tyulbashev2018b}), we visually checked the number of pulses visible in known pulsars before and after the neural network operation. It turned out that the neural network removed 6\% of previously detected pulses.

In addition to the known pulsars and RRATs, the network detected pulses that were missed during visual viewing in the earlier search. Some of these missed pulses turned out to be interferences, some were pulses of known pulsars recorded in the side lobes, some were presumably new RRATs. Fig. 1a-e shows examples of interferences that the neural network considered transients.

   \begin{table*}
	\centering
	\caption{Characteristics of the found RRATs.}
	\label{tab:table1}
	\begin{tabular}{lccccccccc} 
		\hline
	name&$\alpha_{2000}$&$\delta_{2000}$&$DM$~(pc/cm$^3$)&$W_{0.5}$~(ms)&$S/N$&$S_{peak}$~(Jy)& $S_{int}$~(mJy)&$N_1/N_2$  &$n$~(1/hour)\\
	\hline	

J0034+27&$00^h34^m57^s$&$27^o47^\prime$&16&18&7.0&2.4&<0.17&1/2&0.017\\
J1346+06&$13^h46^m11^s$&$06^o10^\prime$&9&22&12.6&5.8&<0.23&1/10&0.085\\
J1432+09&$14^h32^m30^s$&$09^o08^\prime$&14&40&20.0&8.6&<0.22&1/2&0.017\\
J2002+13&$20^h02^m07^s$&$13^o03^\prime$&5&22&8.0&4.8&<0.30&1/3&	0.026\\
		\hline
	\end{tabular}
\end{table*}

  \begin{figure}
   \centering
  \includegraphics[width=0.9\columnwidth]{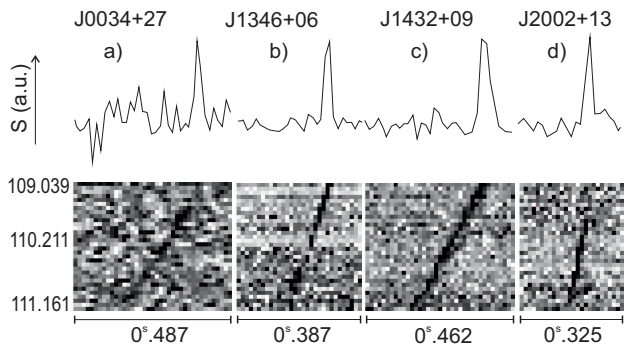}
   \caption{Dynamic spectra and pulse profiles of new RRATs are on the figure. The axes are the same with Fig.~1.}
    \end{figure}
    
During the visual check, the candidates selected by the neural network, shown in Fig.1a-e, would have been eliminated. Let's consider the reasons for screening out these candidates. To obtain an estimate of $S/N$, it is necessary to calculate $\sigma_n$. The processing program calculated for each beam $\sigma_n$ at an interval of one hour, after equalizing the gain in the frequency channels, adding them on $DM=0$~pc/cm$^3$ and removing pulse interference. For the profile in Fig.1a, it turned out that $S/N \approx 5$. In the published papers, the ratio $S/N \ge 7$ was used for the RRATs found. Candidates up to $S/N=5$ were recorded in the search program. Practical work has shown that at $S/N=5-6$, an inclined line of dark pixels is still visible on the dynamic spectrum, distinguishing a RRAT candidate from interference. If we were detecting a RRAT candidate with $S/N\ge 7$, then the weaker pulses observed in the candidate's direction were used to confirm it and search for a period. For the case considered in Fig.1a, the RMS deviations of the noise on the profile were apparently slightly higher than the value determined on the hourly interval, and yet, you can see an inclined line in the drawing. We cannot say unequivocally that we observe an interference, but the $S/N$ estimate is still less than the $S/N>6-7$ criterion generally accepted in radio astronomy, which does not allow us to publish such RRATs as new open sources without additional good reasons. There are $\sim 500-700$ candidates remaining with $S/N<6$ and a visually detectable dispersion delay line after the neural network processed all the images, and we reject them as unreliable. At the same time, the fact that the neural network finds candidates that visually have a weakly pronounced line in the dynamic spectrum, indicates its reliable operation. We are sure that some of these candidates with $S/N<6$ are new RRATs and we are developing ways to confirm them.

Fig. 1b shows the case of an interference located in the frequency domain. The processing program was unable to remove this frequency interference. Fig. 1c shows interference that was simultaneously observed in both the frequency and time domains. Fig. 1d shows a case of poor subtraction of the baseline by the processing program. The horizontal line on the profile shows from which level the search program determined $S/N$. According to the formal feature, the transient is found, but, firstly, the real $S/N <5$, and this is evident by the size of the noise track. Secondly, the line showing the dispersion signal is not visually observed. This indicates that the signal profile is obtained from a small number of pixels, and it indirectly indicates a false detection of the transient. Thirdly, the observed profile consists of a single point. On the supposed $DM=85$~pc/cm$^3$, the minimum width of the profile due to dispersion smoothing inside the frequency channel (the channel width is 78 kHz, the central frequency of observations is 110.25 MHz) should be equal to 4 points. With visual inspection, such candidates are easily excluded. The neural network left this candidate as RRAT, having ''seen``, apparently, several bright pixels lined up along the red line. Fig. 1e shows the type of interference missed during neural network training. The nature of this interference is not clear. It is easily eliminated during visual viewing, and we hope to get rid of this kind of interference with the next iteration of the neural network. The total number of this kind of interference is not large compared to the total number of interferences. When processing semi-annual data, several thousand of them were recorded.

The neural network also found new pulses of known strong pulsars observed in the side lobes of the LPA antenna. These pulses were detected by the network once, and earlier during our visual search (\citeauthor{Tyulbashev2018b}, \citeyear{Tyulbashev2018b}) were missed. Generally speaking, the leakage of sources into the side lobes of the LPA LPI is a serious problem. For example, Fig. 2 in the paper \citeauthor{Tyulbashev2021a} (\citeyear{Tyulbashev2021a}) shows the detected pulses of known pulsars located outside the boundaries of the investigated area. The figure shows that in the studied area with declination $+56^o<\delta <+87^o$ pulsars in the side lobes in different directions in the sky were observed more often than pulsars caught in the main lobe of the antenna. Detection of previously undetected pulses of real pulsars in the side lobes of the antenna again indicates a higher efficiency of the neural network compared to the human eye.

In addition to interferences and pulses of known pulsars, the neural network found objects that could not be identified with either known pulsars and RRATs, or with interference (Fig.~2). The found pulses were detected in one beam and their $S/N>7$. Table 1 shows the characteristics of the detected pulses. In the first column, there is the name of a source, in the second and third columns there are coordinates for the year 2000. The declination error for all found sources is defined as half the distance between adjacent rays and is equal to $\pm 15^\prime$. The error in right ascension is defined as the size of the receiving beam of the LPA at half power and is equal to $\pm 1.5$m. In columns from the fourth to the seventh, $DM$ pulses, half-widths of profiles ($W_{0.5}$), $S/N$ of strongest pulses, peak flux density ($S_{peak}$) are given. In the eighth column, an upper estimate of the integral flux density ($S_{int}$) is given. The search for the regular pulsar emission in the direction of the found RRAT to obtain an estimate of $S_{int}$, was carried out using summed power spectra and summed periodograms (\citeauthor{Tyulbashev2022}, \citeyear{Tyulbashev2022}). Monitoring data was used for the search, for which the full frequency band was 2.5 MHz, the frequency channel width was 78 kHz, and the sampling time of one point was 12.5 ms. The absence of signals with a known $DM$ in the summed power spectra allowed us to give upper estimates of $S_{int}$. $S_{peak}$ estimates and upper estimates of  $S_{int}$ may be underestimated, since the coordinates of the assumed RRATs are known with a large error, and we cannot make corrections to the flux density that take into account the possible divergence of the direction to the center of the LPA beam and the direction to the source. The values given in columns 7 and 8 can be up to 1.5-2 times higher. Sensitivity, when searching for ordinary pulsar radiation, depends on many factors: frequency channel width, pulse width, pulsar's $DM$, etc. (\citeauthor{Tyulbashev2022}, \citeyear{Tyulbashev2022}). Our upper estimates of the integral flux density are given under the assumption that the pulsar period $P>0.5$~s. 

The ninth column $N_1$ contains information on the number of candidate RRAT pulses found by the neural network when reviewing images of dynamic spectra obtained during the processing of semi-annual observations. To confirm the detected pulses, we processed data from August 2014 to December 2019. $N_2$ in the ninth column is the total number of pulses found during a special search over an interval of 5.5 years. The pulse search scheme was standard: subtraction of the baseline; addition of frequency channels without and taking into account $DM$; calculation of standard deviations and comparison of the observed peaks with noise; checking the signal height before and after $DM$ compensation; visual control of all obtained dynamic spectra (more about processing in \citeauthor{Brylyakova2021} (\citeyear{Brylyakova2021})). The search was conducted up to $S/N\ge 5$. The level $S/N \ge 6$ was considered a reliable detection subject to visual detection of a line having a dispersion shift in frequency channels in the dynamic spectrum. The tenth column shows the frequency of occurrence of pulses ($n$), taking into account the new pulses found during their purposeful search over an interval of 5.5 years. 
    
Comments on candidates for new RRATs after searching by known coordinates and a known $DM$ on the interval of 5.5 years:\\
\textbf{J0034+27}. 6 pulses with a pronounced line of dispersion delays were detected for the transient. Out of these 6 pulses, 2 have $S/N\ge 6$ on the interval 5.5 years. Taking into account the passage of the source through the LPA receiving beam at half power during 3-4 minutes, one pulse with a detection threshold $S/N\ge 6$ from this RRAT comes in a time equivalent to 50-60 hours of continuous observations;\\
\textbf{J1346+06}. 26 pulses were found for the transient, of which 10 pulses with $S/N \ge 6$. The three pulses have a $S/N>10$;\\
\textbf{J1432+09}. 8 pulses were found for the transient, of which 2 pulses with $S/N \ge 6$;\\ 
\textbf{J2002+13}. 17 pulses were found for the transient, of which 3 pulses with $S/N \ge 6$.

\section{Discussion of results and conclusion}

The neural network has found single pulses that were missed during visual search due to the method used of fast data viewing. The half-widths of the profiles of the found pulses are characteristic for ordinary seconds duration pulsars. The practical application of the network shows that 6\% of the real pulses of all previously found pulses were missed, and 1.3\% of false sources that are interference were detected. There are a few interferences remaining, and therefore their visual inspection is nota problem. If before the operation of the neural network, 3/4 of the candidates being tested were interferences, then after the operation of the network, the number of interferences decreased by 80 times before a few thousand.

In any case, when a dispersed signal is detected, a new pulse search is performed on all data in the direction of the found source and assuming a known $DM$. If the source had impulses ''thrown out`` by the network, then they will be detected again upon repeated search. Thus, the real losses will include pulses that were detected once during the entire search time interval.

For all detected sources of pulsed emission, the average time between pulses is in the range of 10-60 hours. Earlier in \citeauthor{Logvinenko2020} (\citeyear{Logvinenko2020}), RRATs have already been noted, in which the average time between pulses exceeded 10 hours. It is known that the appearance of RRAT pulses can obey the Poisson distribution (\citeauthor{Meyers2019}, \citeyear{Meyers2019}). In the paper \citeauthor{Smirnova2022} (\citeyear{Smirnova2022}) it was shown that the distribution of nullings (i.e., the duration of intervals without pulses) for some RRATs obeys an exponential law. To obtain a correct estimate of the characteristic time between pulses, it is necessary to know the distribution law. However, we have detected such a small number of pulses that we cannot say anything about their distribution. Therefore, it is impossible to give an unambiguous answer about the existence of a RRAT sample with very long arrival times between successive pulses. We only note that our observations give indications on a possibility of such RRAT samples existence.

At the present time 45 pulsars were discovered in PRAO\footnote{$https://bsa-analytics.prao.ru/en/transients/rrat/$} which in original papers \citeauthor{Shitov2009} (\citeyear{Shitov2009}), \citeauthor{Tyulbashev2018a} (\citeyear{Tyulbashev2018a}), \citeauthor{Tyulbashev2018b} (\citeyear{Tyulbashev2018b}), \citeauthor{Logvinenko2020} (\citeyear{Logvinenko2020}, \citeauthor{Tyulbashev2021a} (\citeyear{Tyulbashev2021a}), \citeauthor{Samodurov2022} (\citeyear{Samodurov2022}), and this paper are called rotating transients. After the discovery of these RRATs by sporadically appearing pulses, some of them were found to have ordinary pulsar radiation. A natural question arises about what is a rotating transient, and how does it differ from an ordinary pulsar?

In the Introduction to this paper, we have already written about the probability that RRATs, like ordinary pulsars, may have different natures and the same external manifestations in the form of irregularly appearing pulses. According to the paper  \citeauthor{Weltevrede2006} (\citeyear{Weltevrede2006}), RRATs can be ordinary pulsars with an unusually long tail in the histogram of the distribution of pulses by energy. In this case, the pulsar (regular) emission may be too weak to register in the observation session, but strong pulses from the tail of the distribution can be detected. In the paper \citeauthor{Tyulbashev2021a} (\citeyear{Tyulbashev2021a}), eight known seconds duration pulsars are given (J0141+6009; J0157+6212; J0653+8051; J1059+6459; J1840+5640; J1910+5655; J2337+6151; J2354+6155) which, in observations on the LPA LPI radio telescope, will look like RRAT, if they are removed to such a distance that with the usual addition of pulsar pulses with a known period, they are not detected in one observation session. In the paper, the search for dispersed pulses on declinations $+56^o<\delta < +87^o$  was carried out, and the viewing area was 4,000 sq.deg. Consequently, when viewing the entire sky at the LPA LPI, one would expect the detection of $\sim 80$ such pulsars, with a uniform distribution of pulsars in the sky.

An additional example indirectly supporting the hypothesis of (\citeauthor{Weltevrede2006}, \citeyear{Weltevrede2006}), are surveys on the search for dispersed pulses conducted at close frequencies (111 and 135 MHz; (\citeauthor{Tyulbashev2018b}, \citeyear{Tyulbashev2018b}, \citeauthor{Sanidas2019}, \citeyear{Sanidas2019}). Some of the sources (J0317+13, J1404+11, J1848+15, J2051+12, J2209+22), determined as RRAT in the Pushchino survey, were detected on LOFAR as ordinary pulsars with separate strong pulses. Such situation developed due to the fact that in the two-hour observation session of the LOTAAS survey, the sensitivity of observations was 2-4 times higher than in the 3.5-minute observation session in Pushchino multibeam pulsar search (PUMPS) (\citeauthor{Tyulbashev2022}, \citeyear{Tyulbashev2022}), which made it possible to register Pushchino RRAT as ordinary seconds duration pulsars. Therefore, there is no doubt that the \citeauthor{Weltevrede2006} (\citeyear{Weltevrede2006})  hypothesis is valid for some part of the RRATs.

Another hypothesis was proposed in the paper \citeauthor{Zhang2007} (\citeyear{Zhang2007}). According to this paper, RRATs are an extreme case of nulling. Pulsars with nullings are known since 1970 (\citeauthor{Backer1970}, \citeyear{Backer1970}). The degree of pulsar nulling varies widely from cases when individual pulses are skipped to cases when the degree of skipping can reach 93\% of all expected pulses (Table 2.1, \citeauthor{Gajjar2014} (\citeyear{Gajjar2014})). Nulling has been actively investigated in the decimeter wavelength range ((\citeauthor{Gajjar2014} (\citeyear{Gajjar2014}) and references there), but it is also observed in the meter range. For example, for the pulsar J0810+37 opened in Pushchino (\citeauthor{Tyulbashev2017}, \citeyear{Tyulbashev2017}), the degree of nulling within a 3.5 minute observation session is in the range of 10-90\% in different observation sessions, with the average degree of nulling 40\%  (\citeauthor{Teplykh2019}, \citeyear{Teplykh2019}, \citeauthor{Teplykh2022}, \citeyear{Teplykh2022}). At the same time, there are cases when not a single pulse was registered for several days in a row in the conducted observation sessions.

We were not able to find any papers with a special study of nulling of RRAT pulsars. However, many papers give an estimate of the number of observed pulses per hour of time, and if the RRAT period is known, then you can roughly estimate the degree of nulling. Using estimates of the pulse arrival rate and period for 21 RRATs (J0139+33; J0848-43; J1005+30; J1129-53; J1132+25; J1226-32; J1317-5759; J1336+33; J1443-60; J1502+28; J1654-23; J1745-30; J1753-38; J1753-12; J1819-1458; J1826-14; J1839-01; J1846-02; J1848-12; J1913+1333; J2105+19), for which the usual pulsar emission was not detected (\citeauthor{McLaughlin2006}, \citeyear{McLaughlin2006}, \citeauthor{Burke-Spolaor2010}, \citeyear{Burke-Spolaor2010}, \citeauthor{Brylyakova2021} \citeyear{Brylyakova2021}, \citeauthor{Smirnova2022}, \citeyear{Smirnova2022}), we have made estimates of nulling. It turned out that for these RRATs, nulling is in the range of 92.96-99.99\%, where the minimum nulling of 92.96\% was observed for the source J1226-32, the maximum nulling is observed for the source J1745-30, and the median value is 99.86\% for the source J1846-02. That is, the nulling values are noticeably higher than those of known pulsars with nullings (\citeauthor{Gajjar2014}, \citeyear{Gajjar2014}) and the assumption of possible extreme nullings of RRAT pulsars (\citeauthor{Zhang2007}, \citeyear{Zhang2007}) may be applicable for a part of the pulsar RRAT sample. On the other hand, the degree of nulling will be determined reliably in the case when we can guarantee that we see \textbf{all} the emitted pulses, and this depends on an antenna sensitivity.

We have already considered pulsars above, in which regular radiation cannot be found in one observation session, while individual pulses from the tail of the energy distribution of pulses are visible. Such pulsars with a long tail of the energy distribution of pulses can be confused with nulling pulsars. However, if the sensitivity of an antenna increases to higher and higher levels, more and more pulses from the tail of the distribution should be observed for such pulsars. All our observations are conducted on the same LPA antenna and at the same sensitivity. Therefore, it is impossible to directly check whether the studied RRAT have weaker pulses. 


However, if a weak regular emission does exist, then we can detect it accumulating a periodic signal. The search for regular pulsar emission showed that with detectable pulses in the dozens of Jy, for some of these RRAT, pulsar emission was not detected at a frequency of 111 MHz in the summed power spectra and summed periodograms. The absence of pulsar emission detection makes it possible to give an upper estimate of the integral flux density of non-nulling pulsars at the level of $S_{int}<0.2-0.3$~mJy  (\citeauthor{Tyulbashev2022}, \citeyear{Tyulbashev2022}). In our opinion, for such pulsars, the hypothesis that RRAT can be pulsars with extreme nullings is supported by observations in the meter wavelength range.      

The hypothesis of a possible manifestation of pulsar activity in the form of emitted giant pulses is supported for five RRATs (J0139+33; J0640+07; J1005+30; J1132+25; J1336+33), discovered on LPA LPI (\citeauthor{Brylyakova2021} (\citeyear{Brylyakova2021}), \citeauthor{Tyulbashev2021} (\citeyear{Tyulbashev2021})). For these RRATs, it is shown that the observed pulses are at least 30 times higher than the expected or observed peak flux densities in the average profiles, and at the same time, the energy distribution of the pulses is described by power-law or lognormal law with a power-law tail, which is typical for pulsars with giant pulses.

Study of 16 sources (\citeauthor{Brylyakova2021} (\citeyear{Brylyakova2021}), \citeauthor{Tyulbashev2021} (\citeyear{Tyulbashev2021}), \citeauthor{Smirnova2022} (\citeyear{Smirnova2022})), discovered as RRATs (\citeauthor{Tyulbashev2018a} (\citeyear{Tyulbashev2018a}),  \citeauthor{Tyulbashev2018b} (\citeyear{Tyulbashev2018b})), for which at least 90 pulses were detected over 5.5 years of daily observations showed that they do not exhibit special properties compared to ordinary, albeit rarely encountered pulsars. It is possible that the fundamental difference with conventional pulsars will be revealed in the $P/\dot P$ diagram. 

Finally, in observations in meter wavelength range, as well as in observations in the decimeter wavelength range, hypotheses are confirmed that part of the RRATs are pulsars with a long tail of the energy distribution of pulses, and part of the RRATs cannot be distinguished from pulsars with long nullings. It was also shown that part of RRATs  is apparently pulsars with giant pulses. 

During the search for new RRATs on the LPA LPI, transients with pulses that appear more and more rarely in time are detected. In this paper, two RRATs were detected (J0032+27; J1434+09), which have the average time between pulses having $S/N \ge 6$, reaches approximately 60 hours. 

\section{Acknowledgment}
The authors thank E.A. Isaev for his help in organizing the work and L.B. Potapova for her help in execution of the paper.


\end{document}